\documentclass[english,twocolumn,aps,pra,longbibliography,crop,tikz,superscriptaddress,nobalancelastpage]{revtex4-2}
\usepackage[english]{babel}
\usepackage[utf8]{inputenc}
\usepackage{amsfonts}
\usepackage{amsmath}
\usepackage{afterpage}
\usepackage{graphicx}
\usepackage{listings}
\usepackage{xcolor}  % For color
\newcommand{\req}[1]{\eqref{#1}} 
\newcommand{\rs}[1]{section~\ref{#1}} 
 
\newcommand{\rf}[1]{Fig.\,\ref{#1}} 
\newcommand{\rt}[1]{table\,\ref{#1}} 

\newcommand{\red}[1]{{\color{red} #1}}
\usepackage{bbold}
\graphicspath{ {./images/} }
\usepackage[colorinlistoftodos, color=green!40, prependcaption]{todonotes}
\usepackage{amsthm}
\usepackage{mathtools}
\usepackage{physics}
\usepackage{xcolor}
\usepackage{graphicx}
\usepackage[left=23mm,right=13mm,top=35mm,columnsep=15pt]{geometry} 
\usepackage{adjustbox}
\usepackage{placeins}
\usepackage[T1]{fontenc}
\usepackage{lipsum}
\usepackage{csquotes}
\usepackage[pdftex, pdftitle={Article}, pdfauthor={Author}]{hyperref} % For hyperlinks in the PDF
\begin{document}
\title{Analysis of polymerized superconducting circuits}

\author{Sean T. Crowe}
\thanks{Corresponding Author: sean.t.crowe2.civ@us.navy.mil}
\affiliation{Naval Information Warfare Center Pacific, San Diego, CA, 92152, United States}
\author{Stefan Evans}
\affiliation{Naval Information Warfare Center Pacific, San Diego, CA, 92152, United States}
\author{Alexei Smolyaninov}
\affiliation{Naval Information Warfare Center Pacific, San Diego, CA, 92152, United States}

\begin{abstract}
We apply polymer quantization, a quantization technique sometimes used in high energy physics, to several superconducting circuits including: transmons, transmission line resonators, and LC circuits. In the case of transmon qubits and transmission line resonators, experimental predictions are very close to what is found with canonical quantization, though in this approach constant charge offsets can be interpreted as quantization ambiguities. In the case of LC circuits, polymer quantization predicts nonlinearities which are not present in the canonical approach. Based on this analysis we design and analyze a qubit which uses a meander inductor instead of a Josephson junction. Implications for qubit performance and fabrication are discussed. Given a choice for an effective phase operator, relevant parameters such as anharmonicity, frequency, and dispersive shifts are calculated for this meander inductor based qubit.
\end{abstract}


\maketitle

\section{Introduction} \label{sec:intro}

Superconducting circuits have emerged as one of the leading platforms for quantum computation and simulation, owing to their scalability, flexible design, and compatibility with existing microwave technologies. Among these, qubits such as the Cooper-pair box, transmon, and flux qubit are modeled using continuous-variable degrees of freedom—namely, the order parameter phase difference across a Josephson junction and its conjugate charge. These systems are typically quantized using the canonical quantization, implying a continuous and separable Hilbert space structure.

Top down approaches for quantization of these systems consist of selecting a lumped element circuit to model the circuit, constructing a classical Hamiltonian system to generate the circuits dynamics, and then proceeding with canonical quantization \cite{Krantz_2019}. Ambiguities arise because it is not always clear what lumped element circuit is appropriate to model the layout. Additionally, depending on the lumped element model, quantization of constrained systems must be used which is technically more challenging \cite{dirac_bracket}. Josephson harmonics in the junctions lead to ambiguities in the transfer function \cite{Harmonics}. Also, as usual, canonical quantization is ambiguous because representations are not generally singled out apriori, the variables used for canonical quantization are not always known, and because ordering ambiguities may be present.  These modeling ambiguities generally lead to physically different predictions which must be resolved by experiment. 

To make matters worse, it is not clear whether or not the application of canonical quantization to lumped element Hamiltonians with the charge basis is the right approach to modeling superconducting circuits. Application of the charge basis to realizable systems like a transmon shunted by an inductor leads to contradictions implying the need for a larger Hilbert space or modified Hamiltonians \cite{stace2020}. Bottom up approaches have also been developed \cite{stace2025}, where charge and phase operators, and their modified algebra are derived from the underlying BCS theory, but to our knowledge these approaches remain experimentally untested.

Polymer quantization \cite{Corichi,Ashtekar,Bojowald,Can_grav}, a non-standard representation of the canonical commutation relations inspired by loop quantum gravity, incorporates a fundamentally discrete basis for the Hilbert space. This framework does not admit well-defined operators for canonically conjugate variables simultaneously, in compliance with the Stone–von Neumann theorem, and is naturally suited for systems where one observable—such as charge—is inherently discrete.

In this work, we explore the application of polymer quantum mechanics to superconducting circuits, focusing on charge-based implementations such as the transmon. By taking the charge operator $\hat{n}$ as fundamental and discrete, such that: $\hat{n}|n\rangle=n |n\rangle$, where $n\in \mathbb{Z}$, and approximating the conjugate phase variable $\hat{\varphi}$ with finite translation operators or equivalently periodic functions, we reformulate superconducting circuit Hamiltonians within a polymer framework. The resulting theory is consistent with existing results for transmon qubits and does not modify the energy level structure. However, in this framework the charge offset can be interpreted as a quantization ambiguity which labels unitarily inequivalent representations. Taking the charge basis as fundamental, we also apply polymer quantization to ordinary LC oscillators and transmission line resonators. For these applications, we do find departures from the canonical quantization which leads to nonlinearities and modifications to the energy level structure. 

An exploration of different quantization techniques for superconducting circuits is relevant because these theoretical considerations inform circuit designs and modalities. For example, in the polymer framework, its possible that Josephson junctions can be replaced with more robust inductive elements without sacrificing anharmonicity.
Josephson Junction (JJ) uniformity is critically important in superconducting electronics and, especially, in the fabrication of qubits. Variation in junction parameters directly translates into significant differences in qubit frequencies across a processor, making it difficult to control and address individual qubits precisely. JJ non-uniformity introduces frequency crowding and reduces the “tunability” space, making it difficult to avoid cross talk. This affects gate fidelity, increases error rates, and complicates the calibration and addressing of each qubit. Likewise, typical quantum error correction protocols require a large number of qubits with uniform properties. Most critically, even small inconsistencies in junction dimensions and material interfaces can introduce decoherence mechanisms, degrading qubit performance and reducing their coherence time \cite{kjaergaard2020superconducting,gambetta2017addressing,mukhanov2011energy}. A qubit based on inductive elements, therefore, may circumvent many of these issues in addition to the obvious advantages in lithography of a simplified material stack owing to the reduced need for a barrier layer during fabrication. This lithographically simplified qubit design could facilitate higher qubit density in a 3D-stacked multilayer structure relative to traditional JJ implementations that rely on an insulating oxide barrier layer, for example. 

This article is organized as follows. In \rs{sec:Approach}, we derive the classical Hamiltonian systems describing: superconducting transmons, transmission line resonators, and LC oscillators. These classical theories are then promoted to quantum theories using polymer quantization, and their characteristics are analyzed. Additionally, ambiguities associated with this approach, such as the choice of effective canonical variables, and quantization ambiguities are discussed. In \rs{sec:NA}, we optimize nonlinear features predicted from polymer quantum mechanics in order to develop a workable design for a superconducting qubit based on a capacitor shunted by a meander inductor. This design is then analyzed using analytic and numerical methods to predict its behavior.

      %I believe leaving the sections in separate files is more organized, change it if you desire -- agreed!
\section{Polymer quantization of superconducting circuits}\label{sec:Approach}

\subsection{Superconducting qubits}\label{polymerQubit}
The first example of a superconducting  qubit is a fixed frequency qubit, which consists of a capacitor in series with a Josephson junction. This type of qubit is modeled by the circuit diagram in~\rf{fig:basic_qubit}. Here, the Josephson junction is characterized by the Josephson inductance, $L_0$. Ideal Josephson junctions are modeled as non-linear inductors and the voltage is given by: 
\begin{equation}\label{V_j}
    V_j=-\frac{L_0}{\sqrt{1-\left(\frac{I}{I_c}\right)^2}}\frac{dI}{dt},
\end{equation}
\begin{figure}
    \centering
    \includegraphics[width=0.2\textwidth]{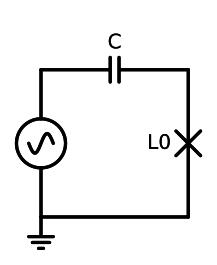}
    \caption{A fixed frequency superconducting qubit, consisting of a capacitor in series with a Josephson junction.}
    \label{fig:basic_qubit}
\end{figure}
Where $L_0=\frac{\Phi_0}{2 \pi I_c}$, $\Phi_0=h/(2e)$, and $I_c$ is the critical current of the junction. Using \eqref{V_j}, Kirchhoff's rule, and the voltage across a capacitor, the classical equation of motion for this system is:
\begin{equation}\label{EOM1}
    V_d(t)=\frac{L_0}{\sqrt{1-\left(\frac{I}{I_c}\right)^2}}\frac{dI}{dt}+\frac{Q}{C}.
\end{equation}
From here we integrate to move the square root factor inside the derivative and replace $I=\frac{dQ}{dt}$ to rewrite \eqref{EOM1} as
\begin{equation}\label{EOM_2}
    V_d(t)= \frac{d}{dt}\left(L_0 I_c\arcsin{\left(\frac{\frac{dQ}{dt}}{I_c}\right)}\right)+\frac{Q}{C}.
\end{equation}

We now derive a Hamiltonian, which will generate the equation of motion at \eqref{EOM_2}. This can be done by positing Hamilton's equations and solving for the Hamiltonian and its canonical variables. Hamilton's equations are:
\begin{equation}
    \begin{split}
        \frac{dQ}{dt}&=\frac{\partial H}{\partial P_Q},\\
        \frac{dP_Q}{dt}&=-\frac{\partial H}{\partial Q},\\
    \end{split}
\end{equation}
where $P_Q$ is the canonical momentum of the charge, whose form will determined. By examining \eqref{EOM_2}, we identify:
\begin{equation}\label{ham_1}
    \begin{split}
    P_Q&=L_0 I_c\arcsin{\left(\frac{\frac{dQ}{dt}}{I_c}\right)},\\
    \frac{\partial H}{\partial Q}&=-V_d(t)+Q/C.\\
    \end{split}
\end{equation}
Solving the first equations here for the time derivative of the charge on the capacitor, we find, $\frac{dQ}{dt}=I_c \sin{\left(P_Q/\left(L_0 I_c\right)\right)}$. Because of Hamilton's first equation, this implies
\begin{equation}\label{ham_2}
    \dot{Q}=\frac{\partial H}{\partial P_Q}=I_c \sin{\left(\frac{P_Q}{\Phi_0/(2 \pi)}\right)}.
\end{equation}
Using \eqref{ham_1} and \eqref{ham_2}, we solve for the Hamiltonian function by integrating. Integrating \eqref{ham_2} by $P_q$ gives:
\begin{equation}\label{ham_3}
    H=-I_c^2 L_0 \cos\left(\frac{P_Q}{L_0 I_c}\right)+G(Q),
\end{equation}
where G is function of the charge, which we  determine by inserting \eqref{ham_3} into \eqref{ham_1} we find the constraint on G to be
\begin{equation}
    G'(Q)=-V_d(t)+Q/C.
\end{equation}
The integration is straightforward and the result is,
\begin{equation}
    G(Q)=\frac{1}{2 C}Q^2-V_d(t) Q+E_0.
\end{equation}
Here $E_0$ is a constant of integration, which does not affect the dynamics. We set it to zero for the remainder of our analysis. Putting this all together, we have the Hamiltonian function:
\begin{equation}
    H=-I_c^2 L_0  \cos\left(\frac{P_Q}{L_0 I_c}\right)+\frac{1}{2 C}Q^2-V_d(t) Q.
\end{equation}
Because $Q$ and $P_Q$ are canonically related, they have the canonical Poisson bracket: $\left\{Q,P_Q\right\}=1$. Here we make the change of variables: $Q=2e n$ and $P_Q=L_0 I_c \varphi$, where $n$ is the number of Cooper pairs. These new variables have the Poisson bracket,
\begin{equation}\label{bracket}
    \left\{n,\varphi\right\}=\frac{1}{\hbar},
\end{equation}
and they imply the Hamiltonian,
\begin{equation}\label{penultimate_ham}
    H=-I_c^2 L_0 \cos\left(\varphi\right)+4\frac{e^2}{2 C}n^2-2 eV_d(t) n.
\end{equation}
We also use the variables: $E_j=I_c^2 L_0$ and $E_c=e^2/2 C$. Finally the classical Hamiltonian system is given by:

\begin{equation}\label{ClassHam}
    \begin{split}
            H&=-E_j\cos\left(\varphi\right)+4 E_c \left(n-\sqrt{\frac{C}{8 E_c}}V_d(t)\right)^2,\\
            \left\{n,\varphi\right\}&=\frac{1}{\hbar},
    \end{split}
\end{equation}
where we have completed the square and ignored a term which is independent of the canonical variables and therefore doesn't affect the dynamics. In this classical picture, the time dependent voltage term corresponds to a charge offset. 

This classical system cannot be quantized in the canonical sense because the variable $n$ should only be supported on the space of square integrable functions over the integers after quantization. That is, after quantization we should have $\hat{n}|n\rangle=n|n\rangle$ where $n\in \mathbb{Z}$. Supposing we take the canonical route where $n$ and $\varphi$ are promoted to operators where $\left[\hat{n},\hat{\varphi}\right]= i$, The Stone-von Neumann Theorem then implies that $\hat{n}$ is supported on the space of square integrable functions over the real numbers, giving the required contradiction. 

A quantization of this system is still possible though by reformulating the classical algebra to be: 
\begin{equation}
    \begin{split}
        \left\{n,\sin(\varphi)\right\}&=\cos(\varphi)/\hbar,\\
        \left\{n,\cos(\varphi)\right\}&=-\sin(\varphi)/\hbar,\\
        \left\{\sin(\varphi),\cos(\varphi)\right\}&=0.\\
    \end{split}
\end{equation}

This algebra isn't canonical, but is the lie algebra of the euclidean group in two dimentions. A quantization of this algebra can be found from its representation theory which was originally analyzed by Isham \cite{isham1983topological}. No unique quantization is available, but the ambiguity associated with quantization in this case can be parameterized by a single parameter $\theta$. The quantization of this algebra including ambiguities is known \cite{Can_grav}, and is given by,
\begin{equation}
    \begin{split}
        \hat{n}\psi(\varphi)&=i \frac{\partial\psi(\varphi)}{\partial \varphi},\\
        \widehat{\cos(\varphi)}\psi(\varphi)&=\cos(\varphi)\psi(\varphi),\\
        \widehat{\sin(\varphi)}\psi(\varphi)&=\sin(\varphi)\psi(\varphi),\\
        \psi(\varphi)&\in L^2(S^1).
    \end{split}
\end{equation}
This representation amounts to the polymer quantization of this algebra of observables. In this sense, only functions of $n, \cos(\varphi),$ and $\sin(\varphi)$, can be represented in the quantum theory. Because the qubit Hamiltonian consists of these functions, it can be represented. For any given $\theta$ there is an eigenbasis of $\hat{n}$ which is  given by: $\psi_{n}^{\theta}(\varphi)=\mathrm{e}^{-i(n+\theta)\varphi}/\sqrt{2 \pi}$, where $n$ is an integer labeling the $n$th eigenstate. The parameter $\theta$ cannot be removed by a unitary transformation and does lead to observable consequences. Considering the expectation value of $n$ in one of its eigenstates, this is equivalent to the expected number of electrons on the capacitor of the qubit:
\begin{equation}
\begin{split}
    \int_{-\pi}^{\pi}d \varphi \psi^{\theta *}\hat{n}\psi^{\theta}&=\frac{1}{2 \pi}\int_{-\pi}^{\pi}d \varphi \,i\,\mathrm{e}^{i(n+\theta)\varphi}\frac{\partial}{\partial \varphi}\mathrm{e}^{-i(n+\theta)\varphi}\\
    &=\left(n+\theta \right).
\end{split}
\end{equation}

In this approach, the number operator's eigenvalues are only determined up to a constant shift of the number of cooper pairs on the capacitor. This shift has already been identified experimentally, and in the literature is called the charge offset. In particle physics it is called the spin of the algebra. It is thought to be determined by noise stemming from the control lines, which interact with the capacitor part of the qubit or by two level systems in the Josephson junctions \cite{chargeoffset,Roth:2021lrz}. By making the capacitor large and in turn making  $E_c$ small the effect of this uncontrolled parameter is mitigated. The full quantization of \eqref{ClassHam} is then given by:

\begin{equation}
    \boxed{
    \begin{split}
        \hat{H}=-E_j\widehat{\cos\left(\varphi\right)}&+4 E_c \left(\hat{n}+\theta-\sqrt{\frac{C}{8 E_c}}V_d(t)\right)^2,\\
        \left[\hat{n},\widehat{\sin(\varphi)}\right]&=i\widehat{\cos(\varphi)},\\
        \left[\hat{n},\widehat{\cos(\varphi)}\right]&=-i\widehat{\sin(\varphi)},\\
        \left[\widehat{\sin(\varphi)},\widehat{\cos(\varphi)}\right]&=0,\\
        \hat{n}|n\rangle&=n |n\rangle .
    \end{split}}
\end{equation}

\subsection{LC circuits}\label{sec:LCCircuit}

In this section we derive the classical Hamiltonian of an LC circuit and promote it to the corresponding quantum theory using polymer quantization. The classical theory is easier in this case. To start we consider a circuit similar to that in \rf{fig:basic_qubit}, except with the junction replaced by an inductor. To derive the equations of motion we use the Kirchoff voltage law. The voltage across the inductor is $V_L=-L \frac{dI}{dt}$, and the voltage across the capacitor is $V_c=Q/C$, so the equation of motion is:
\begin{equation}
    \begin{split}
        V_d(t)&=V_c+V_L\\
        &=L \frac{dI}{dt}+Q/C\\
        &=\frac{d \Phi}{dt}+Q/C.
    \end{split}
\end{equation}
We also identify: $\dot{Q}=I=\Phi/L$. From here we postulate Hamilton's equations and solve for the Hamiltonian. That is:
\begin{equation}
    \begin{split}
        \dot{Q}&=\frac{\partial H}{\partial \Phi},\\
        \dot{\Phi}&=-\frac{\partial H}{\partial Q}.
    \end{split}
\end{equation}
This then implies that $\frac{\partial H}{\partial \Phi}=\frac{\Phi}{L}$, which gives $H=\frac{\Phi^2}{2 L}+G(Q)$, where G is a function that depends on $Q$. To determine this function, we plug this candidate Hamiltonian into Hamilton's second equation and solve. The result is  $\dot{\Phi}=-G'(Q)=-Q/C+V_d(t)$. This implies that $G=\frac{Q^2}{2 C}-V_d(t)Q$. Finally the Hamiltonian is given by: $H=\frac{\Phi^2}{2 L}+\frac{Q^2}{2 C}-Q V_d(t)$, and the Poisson bracket is $\left\{Q,\Phi\right\}=1$. To make contact with the qubit Hamiltonian from earlier we make the extended canonical transformation:
\begin{equation}
    \begin{split}
        n&=\frac{Q}{2 e},\\
        \varphi&=\frac{2 e}{\hbar}\Phi.
    \end{split}
\end{equation}
The Poisson bracket with these new variables is: $\left\{n,\varphi\right\}=\left\{\frac{Q}{2 e},\frac{2 e}{\hbar}\Phi\right\}=\frac{1}{\hbar}$. Finally the full classical system is given by: 
\begin{equation}\label{ClassHamLC}
    \begin{split}
        H&= \frac{1}{2}E_l \, \varphi^2+4 E_c n^2-\sqrt{2 c E_c} \, n \,V_d(t),\\
        \left\{n,\varphi\right\}&=1/\hbar,
    \end{split}
\end{equation}
where $E_c=e^2/2C$ and $E_l=\hbar^2/(4 e^2 L)$. At the classical level this is a model for a driven harmonic oscillator. 

If we assume in the quantum theory that the observable $\hat{n}$ is only supported on the space of square integrable functions over the integers, then the quantization of this classical system is non-trivial, and cannot be done canonically. Suppose  we have an operator $\hat{\varphi}$ that commutes with $\hat{n}$ according to: $\left[\hat{n},\hat{\varphi}\right]=i$ while $\hat{n}|n\rangle=n|n \rangle$ for $n\in \mathbb{Z}$. Then consider:
\begin{equation}
    \begin{split}
        \hat{n}\exp\left(-i \alpha \hat{\varphi}\right)|m\rangle&=\left(\left[\hat{n},\exp\left(-i \alpha \hat{\varphi}\right)\right]+\exp\left(-i \alpha \hat{\varphi}\right)\hat{n}\right)|m\rangle\\
        &=\left(\alpha \exp\left(-i \alpha \hat{\varphi}\right)+m \exp\left(-i \alpha \hat{\varphi}\right)\right)|m\rangle\\
        &=\left(\alpha+m\right)\exp\left(-i \alpha \hat{\varphi}\right)|m\rangle.
    \end{split}
\end{equation}
Therefore $\exp\left(-i \alpha \hat{\varphi}\right)|m\rangle$ is an eigenstate of $\hat{n}$ with eigenvalue $\alpha+m$,  contradicting the assumption that $\hat{n}$ should have integer valued eigenvalues. We conclude that under these assumptions an operator $
\hat{\varphi}$ does not exist. This makes the canonical quantization of \eqref{ClassHamLC} non-trivial because its canonical variables cannot be directly promoted to operators.

At the kinematical level this system can be quantized in the same way as in \ref{polymerQubit}. The classical algebra is reformulated as: 
\begin{equation}
    \begin{split}
        \left\{n,\sin(\varphi)\right\}&=\cos(\varphi)/\hbar,\\
        \left\{n,\cos(\varphi)\right\}&=-\sin(\varphi)/\hbar,\\
        \left\{\sin(\varphi),\cos(\varphi)\right\}&=0.
    \end{split}
\end{equation}
Then the quantization can be done according to:
\begin{equation}
    \begin{split}
        \hat{n}\psi(\varphi)&=i \frac{\partial\psi(\varphi)}{\partial \varphi},\\
        \widehat{\cos(\varphi)}\psi(\varphi)&=\cos(\varphi)\psi(\varphi),\\
        \widehat{\sin(\varphi)}\psi(\varphi)&=\sin(\varphi)\psi(\varphi),\\
        \psi(\varphi)&\in L^2(S^1).
    \end{split}
\end{equation}
The dynamics of this system are non-trivial because there is no operator available to represent the classical variable $\varphi$. In the polymer approach, what is done to mitigate this issue is to find a suitable approximation of $\hat{\varphi}$ using the periodic functions which can be represented: $\widehat{\sin{\varphi}}$ and $\widehat{\cos \varphi}$. There are however many options. For example, $\widehat{\sin \varphi}\sim \hat{\varphi}$ for low enough energy states. One can also try the junction potential $1-\widehat{\cos \varphi}\sim\frac{1}{2}\hat{\varphi}^2$. Also in light of possible Josephson harmonics which are periodic functions that can be represented \cite{Harmonics}, the number of ways to promote \eqref{ClassHamLC} is very broad. Generally, these dynamics will be anharmonic. 

There are also pathological choices such as $\hat{\varphi}_{eff}=\arcsin{\left(\widehat{\sin(\varphi)}\right)}$. Even choices like these imply some anharmonicity which can be computed using first order perturbation theory. The Hamiltonian is:
\begin{equation}
    \hat{H}= \frac{1}{2}E_l \arcsin{\left(\widehat{\sin(\varphi)}\right)}^2+4 E_c \hat{n}^2.
\end{equation}
To analyze the anharmonicity of this system we make a continuum approximation where $\hat{n}$ is assumed to have eigenvalues in the real line and there is an operator $\hat{\varphi}$ which canonically commutes with $\hat{n}$. The Hamiltonian can then be rearranged into a harmonic term and a perturbative term:
\begin{equation}\label{pertham}
\begin{split}
    \hat{H}&=\frac{1}{2}E_l \hat{\varphi}^2+4 E_c \hat{n}^2+\\
    &\frac{1}{2}E_l \left(\arcsin{\left(\sin(\widehat{\varphi})\right)}^2-\hat{\varphi}^2\right).
\end{split}
\end{equation}
The second term is a perturbation which is not identically zero because arcsin only takes on values in the range $\left[-\pi/2,\pi/2\right]$. The first term is just a normal harmonic term whose eigenstates are up to a normalization factor given by:
\begin{equation}
    \psi_n(\varphi)=N \mathcal{H}_n\left(\sqrt{\sqrt{\frac{E_l}{8 E_c}}}\varphi\right)\exp{\left(-\frac{1}{2}\sqrt{\frac{E_l}{8 E_c}}\varphi^2\right)},
\end{equation}
where $\mathcal{H}_n$ are the Hermite polynomials. Using these eigenstates and the perturbative term of \eqref{pertham} we have numerically computed the anharmonicity of the system according to $\alpha=\left(\Delta_2-\Delta_1\right)-\left(\Delta_1-\Delta_0\right)$, and $\Delta_n=\frac{1}{2}E_l\int d\varphi |\psi_n|^2 \left(\arcsin{\left(\sin(\varphi)\right)}^2-\varphi^2\right)$. The anharmonicity is a function of $E_l/E_c$ which can be computed numerically. We have computed $\alpha/E_c$ for up to $E_l/E_c=100$ and the results are shown in \rf{fig:anharm}.

\begin{figure}
    \centering
    \includegraphics[width=0.45\textwidth]{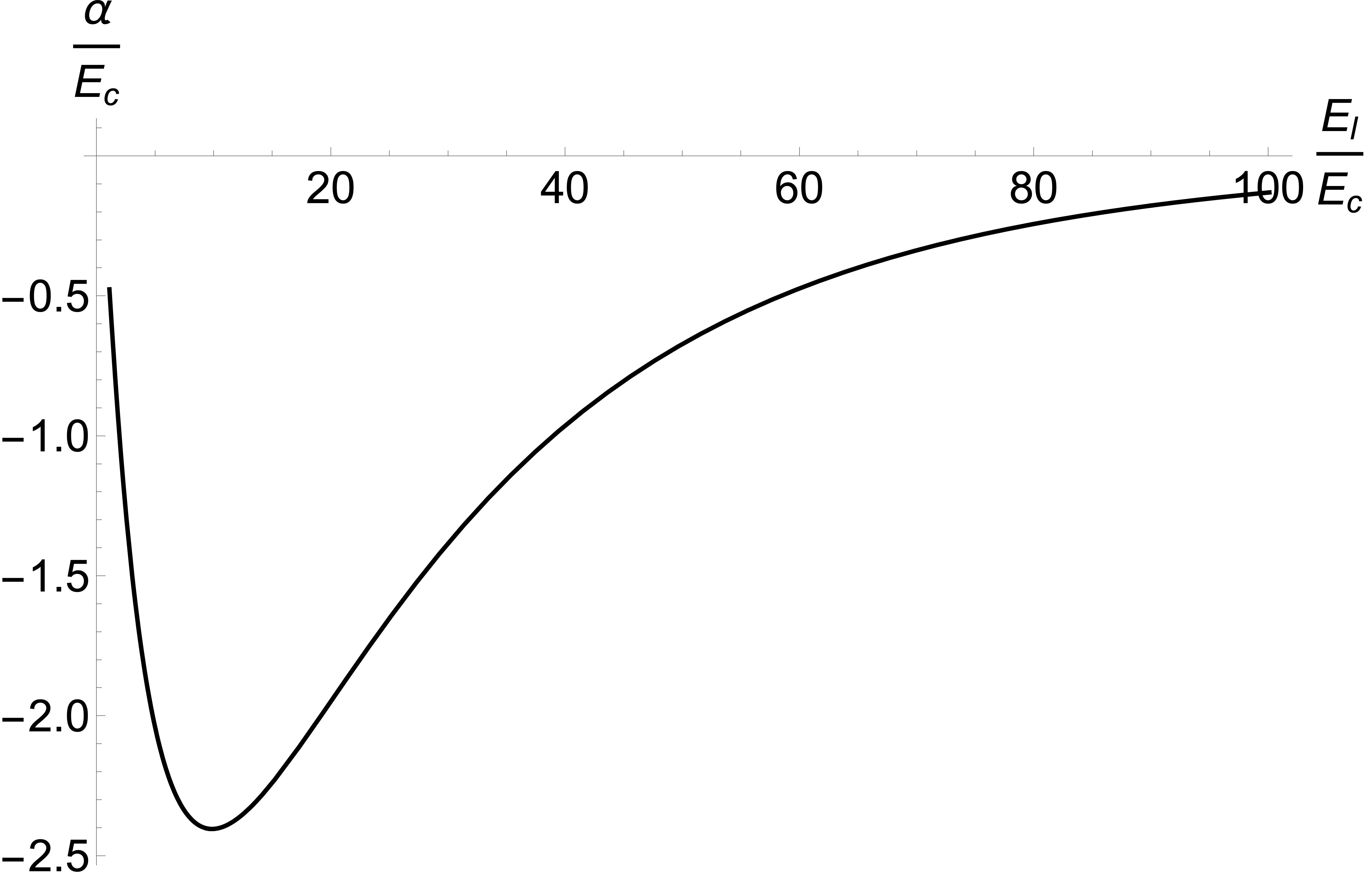}
    \caption{$\alpha/E_c$ vs $E_l/E_c$. As the charge participation ratio increases the anharmonicity decreases exponentially, but there are regions where this choice of $\hat{\varphi}_{eff}$ has higher anharmonicity than the normal Josephson potential.}
    \label{fig:anharm}
\end{figure}

Generally, the polymer quantization of \eqref{ClassHamLC} is then given by:
\begin{equation}
\label{PhiEffAnharm}
    \boxed{
    \begin{split}
        \hat{H}= \frac{1}{2}E_l \, \hat{\varphi}_{eff}^2&+4 E_c \left(\hat{n}+\theta-\sqrt{\frac{C}{8 E_c}}V_d(t)\right)^2,\\
        \left[\hat{n},\widehat{\sin(\varphi)}\right]&=i\widehat{\cos(\varphi)},\\
        \left[\hat{n},\widehat{\cos(\varphi)}\right]&=-i\widehat{\sin(\varphi)},\\
        \left[\widehat{\sin(\varphi)},\widehat{\cos(\varphi)}\right]&=0,\\
        \hat{n}|n\rangle&=n |n\rangle,
    \end{split}}
\end{equation}
where $\hat{\varphi}_{eff}$ is an effective phase variable, which depends on the operators $\widehat{\sin(\varphi)}$ and $\widehat{\cos(\varphi)}$, whose form has to be determined by more detailed analysis of the underlying circuit physics, or by experiment. In \rs{sec:NA}, we design a qubit using the form $\frac{1}{2}\hat{\varphi}_{eff}^2=1-\widehat{\cos{\varphi}}$, which has been sometimes used in cosmological models \cite{Corichi}.

\subsection{Transmission line resonators}

In this section we derive the classical Hamiltonian formulation of a transmission line resonator and quantize it using the polymer approach. The transmission line resonator is modeled by the LC ladder shown in \rf{fig:ladder}. Here we consider the case where $L_i=\Delta L$ and $C_i=\Delta C$. If $N$ is the total number of cells in the transmission line then the inductance and capacitance of the line are given by: $L=N \Delta L$, $C=N \Delta C$. Given some voltage at the start of the resonator and making a loop terminating at the ground below $C_n$, and also a second one terminating at $C_{n+1}$, Kirchoff's equations imply:  
\begin{equation}
\begin{split}
    V(t)&=V_{C_n} + \sum_{i=0}^{i=n}V_{L_i}\\
    &=V_{L_{n+1}}+V_{C_{n+1}}+ \sum_{i=0}^{i=n}V_{L_i},
\end{split}
\end{equation}
where $V(t)$ is the voltage at the start of the resonator. Subtracting the second equation from the first gives:
\begin{equation}\label{K1}
\begin{split}
    0&=V_{L_{n+1}}+V_{C_{n+1}}-V_{C_n}\\
    &=-\Delta L \frac{dI_{L_{n+1}}}{dt}+\frac{1}{ \Delta C}\left(Q_{n+1}-Q_n\right)\\
    &\approx -\Delta L \frac{dI_{L_{n+1}}}{dt}+\frac{1}{\Delta C}\nabla Q_n.
\end{split}
\end{equation}
Also, the Kirchoff current law applied to the node in between $L_n$ and $L_{n+1}$ implies: 
\begin{equation}\label{K2}
    \begin{split}
    I_{L_n}&=I_{L_{n+1}}+I_{c_n}\\
    &\implies\\
    \dot{Q}_n&=I_{L_{n+1}}-I_{L_n}\\
    &\approx \nabla {I_{L_n}}.
    \end{split}
\end{equation}
Taking the gradient of \eqref{K1} and inserting \eqref{K2} gives the wave equation:
\begin{equation}
    0=\ddot{Q}-\frac{1}{\Delta L \Delta C}\nabla^2 Q.
\end{equation}
The Hamiltonian and Poisson bracket generating this equation are given by:
\begin{equation}
    \begin{split}
        H&=\int dn \left(\frac{P_{Q}^2}{2 \Delta L}+\frac{1}{2 \Delta C}\left(\nabla Q\right)^2\right),\\
        \left\{Q,P_{Q}\right\}&=\delta(n-m).
    \end{split}
\end{equation}
From here we make an extended canonical transformation
\begin{equation}
    \begin{split}
        n&=\frac{Q}{2 e},\\
        \varphi&=\frac{2 e}{\hbar}\Phi,
    \end{split}
\end{equation}
which then implies the full classical system:
\begin{equation}
    \begin{split}
        H&=\int dn \left(\frac{1}{2}E_l \, \varphi^2+4 E_c\left(\nabla n\right)^2\right),\\
        \left\{n(n),\varphi(m)\right\}&=\delta(n-m)/\hbar.
    \end{split}
\end{equation}

\begin{figure}
    \centering
    \includegraphics[width=0.45\textwidth]{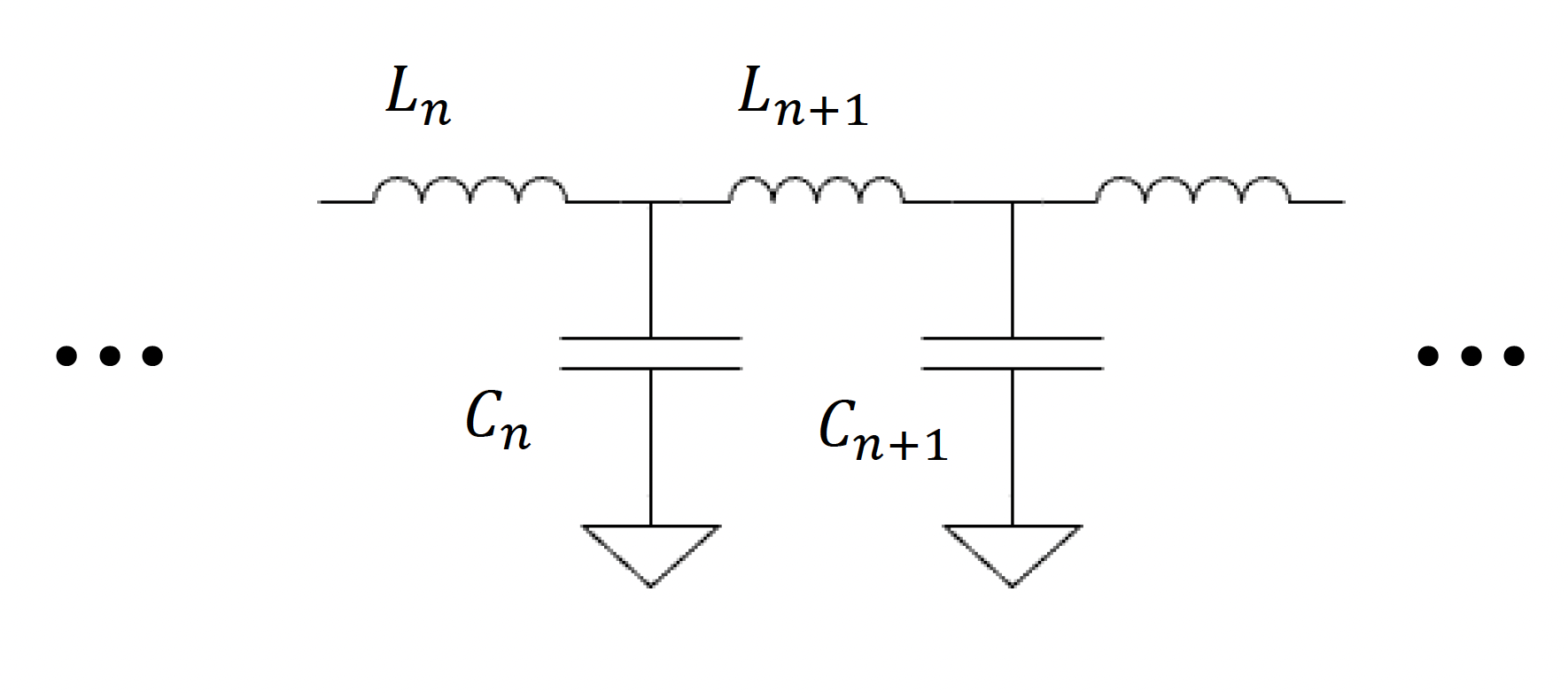}
    \caption{An LC ladder which can be used to model a transmission line resonator.}
    \label{fig:ladder}
\end{figure}

As in the previous section, promotion of this classical system to a quantum system is problematic because no operator to represent $\varphi$ can be found in the case where $\hat{n}$ has a discrete spectrum, which we are assuming here. This system will also have higher order derivative corrections due to the discrete nature of the LC ladder, and we are ignoring these other effects here in order to focus on effects due to the polymer quantization.

This system can be quantized in the same way as in \rs{sec:LCCircuit}, only now with features added because this example is a field theory. The kinematics are given by:
\begin{equation}
    \begin{split}
        \left[\hat{n}(n),\widehat{\sin(\varphi)}(m)\right]&=i\widehat{\cos(\varphi)}\delta(m-n),\\
        \left[\hat{n}(n),\widehat{\cos(\varphi)}(m)\right]&=-i\widehat{\sin(\varphi)}\delta(m-n),\\
        \left[\widehat{\sin(\varphi)}(m),\widehat{\cos(\varphi)}(n)\right]&=0.\\
    \end{split}
\end{equation}

Because $\varphi(m)$ cannot be represented as an operator we choose some operator $\hat{\varphi}_{eff}$, which is a function of $\widehat{\sin \varphi}(m)$ and $\widehat{\cos \varphi}(n)$ to take its place. The final Hamiltonian is: 
\begin{equation}
\label{HamiltonianFinal}
    \begin{split}
        \hat{H}=\int dn \left(\frac{1}{2}E_l \, \hat{\varphi}_{eff}(n)^2+4 E_c \left(\nabla(\hat{n}(n)+\theta(n))\right)^2\right),  
    \end{split}
\end{equation}
where now there is one ambiguity parameter, $\theta$ for each point in the one dimensional transmission line resonator. To make progress on this system we consider the analogous Hamiltonian: 
\begin{equation}\label{TWPAHam}
    \begin{split}
        \hat{H}=\int_{0}^{N} dn \left(\frac{1}{2}E_l \, \hat{\varphi}(n)^2-\frac{1}{4!}E_l \, \hat{\varphi}^4+4 E_c \left(\nabla\hat{n}(n)\right)^2\right),      
    \end{split}
\end{equation}
that corresponds to the case where $\frac{1}{2}\varphi_{eff}^2=1-\cos{\left(\hat{\varphi}\right)}$, and the cosine term as been expand to include terms up to $O(\varphi^4)$. We now make a continuum approximation and assume that $\hat{n}(n)$ takes on eigenvalues in the real line and that $\left[\hat{n}(n),\hat{\varphi}(m)\right]=i \delta(n-m)$. We are also only integrating from 0 to $N$, the number of cells in the transmission line, and we have set $\theta$ to be zero. It is possible to analyze the anharmonicity of the fundamental mode by making a mode decomposition which respects the boundary conditions of a quarter wave resonator:
\begin{equation}
    \begin{split}
        \hat{\tilde{n}}&=\frac{2}{N}\int_0^N dn\sin{\left(\frac{\pi}{2}\frac{n}{N}\right)\hat{n}(n)},\\
        \hat{\tilde{\varphi}}&=\int_0^N dn \sin{\left(\frac{\pi}{2}\frac{n}{N}\right)\hat{\varphi}(n)}.
    \end{split}
\end{equation}

A calculation shows that $\left[\hat{\tilde{n}},\hat{\tilde{\varphi}}\right]=i$, so these mode amplitudes are canonically related. Restricting to just this single mode allows for this relationship to be inverted and the result is:
\begin{equation}
    \begin{split}
        \hat{n}(n)&=\hat{\tilde{n}} \sin{\left(\frac{\pi}{2}\frac{n}{N}\right)},\\
        \hat{\varphi}(n)&=\frac{2}{N}\hat{\tilde{\varphi}} \sin{\left(\frac{\pi}{2}\frac{n}{N}\right)}.\\
    \end{split}
\end{equation}
Inserting these relationships back into the Hamiltonian at \eqref{TWPAHam} and simplifying we find:
\begin{equation}
\begin{split}
    \hat{H}&=\frac{1}{2}E_l \frac{2}{N}\hat{\tilde{\varphi}}^2-\frac{1}{4!}\frac{3}{4}E_l \left(\frac{2}{N}\right)^3
    \hat{\tilde{\varphi}}^4 
    \\
    &\quad+ 4 \left(\frac{\pi}{4} \left(\frac{2}{N}\right)^3 
    E_c\right)\left(\frac{N}{2}\hat{\tilde{n}}\right)^2.
\end{split}
\end{equation}
We make the canonical transformation: $\frac{2}{\sqrt{3}}\hat{\phi}=\frac{2}{N}\hat{\tilde{\varphi}}$ and $\frac{\sqrt{3}}{2}\hat{N}=\frac{N}{2}\hat{\tilde{n}}$, and define the renormalized charging energies: $\tilde{E}_l=\frac{4}{3}\frac{N}{2}E_l$ and $\tilde{E}_c=\frac{3}{4}\frac{\pi}{4}\left(\frac{2}{N}\right)^3E_c$. The resulting Hamiltonian is:
\begin{equation}
\label{HamLine}
    \hat{H}=\frac{1}{2}\tilde{E}_l \hat{\phi}^2-\frac{1}{4!}\tilde{E}_l\hat{\phi}^4+4 \tilde{E}_c  \hat{N}^2.
\end{equation}

Eq.\,\req{HamLine} has the same form as the normal transmon Hamiltonian at this order and standard results can be used. The charge participation ratio for this mode is then given by: $\frac{\tilde{E}_l}{\tilde{E}_c}=\frac{16}{9}\frac{4}{\pi}\left(\frac{N}{2}\right)^3\frac{E_l}{E_c}$. With these renomalized charging energies, typical values for $E_l$ and $E_c$ imply large values of the charge participation ratio of the mode. In \rs{sec:NA}, the readout resonator connecting the qubit to the readout line has a capacitance of $836$ fF, and an inductance of $2.4$ nH. These values imply a charge participation ratio of $E_l/E_c\sim 2960$. In the ideal case where $N\rightarrow\infty$ the renormalized charge participation ratio goes to infinity. In the non-ideal case where $N$ is finite, the charge participation ratio is still very large compared with normal transmons. The anharmonicity of this mode is given by: 
\begin{equation}
\begin{split}
\alpha&=-\tilde{E}_c=-\frac{3}{4}\frac{\pi}{2}\left(\frac{2}{N}\right)^2\frac{e^2}{2 C}.
\end{split}
\end{equation}
When $N\rightarrow\infty$ the anharmonicity of the mode goes to zero, but in the non-ideal case the anharmonicity is small but non-zero. Taking e.g. $N=1000$ (typical for JTWPAs) and using the capacitance value of $836$ fF that we found for the readout resonator, we find the anharmonicity of this fundamental mode to be $\alpha/h=110$ Hz, which is too low to be measured during a typical transmon life time of 100 $\mu$s.

%\input{sections/section02p5}
%\input{sections/section03.tex}
%%%%%%%%%%%%%%%%%%%%%%%%%%%%%%%%%%%%%%%%%%%%%%%%%%%%%%%%%%%%%%%%%%%%%%%%%%%%%%%%%%%%%%%%%%%%

\section{Circuit design and analysis} \label{sec:NA}

\red{
    % Put physical design and analysis here. 
    % Maybe some images of the chip. 
    % What is the inductance of the meander inductor? How does this compare with the hueristic formula? What is the capacitance? What is the resonant frequency of the readout resonator? 
    % What is the coupling strength between the readout resonator and the qubit? What is the predicted anharmonicity of the device? 
    % We should also repeat this analysis for the normal transmon qubit on the other side of the chip. 
    % What goes on in that python notebook?
    }

%%%%%%%%%%%%%%%%%%%%%%%%%%%%%%%%%%%%%%%%%%%%%%%%%%%%%%%%%%%%%%%%%%%%%%%%%%%%%%%%%%%%%%%%%%%%
\subsection{Chip layout}

Taking the above analysis into account, we present and analyze a design here which uses a meander inductor in the place of a Josephson junction. The two-qubit chip in~\rf{FullFig} features a meander inductor based qubit (capacitor in series with a meander inductor) alongside a standard qubit design (capacitor in series with a Josephson Junction). The readout line (1$\leftrightarrow$3) applies an RF signal to scan the resonant frequencies of the readout resonators. This probes the shift in the resonance frequencies while drive lines (2,5) excite the qubits. The flux bias (line 4) applies a magnetic field to the SQUID to control the excitation energy of the transmon qubit.

%%%%%%%%%%%%%%%%%%%%%%%%%%%%%%%%%%%%%
%
\begin{figure}[h]
\center
\includegraphics[width=0.99\columnwidth]{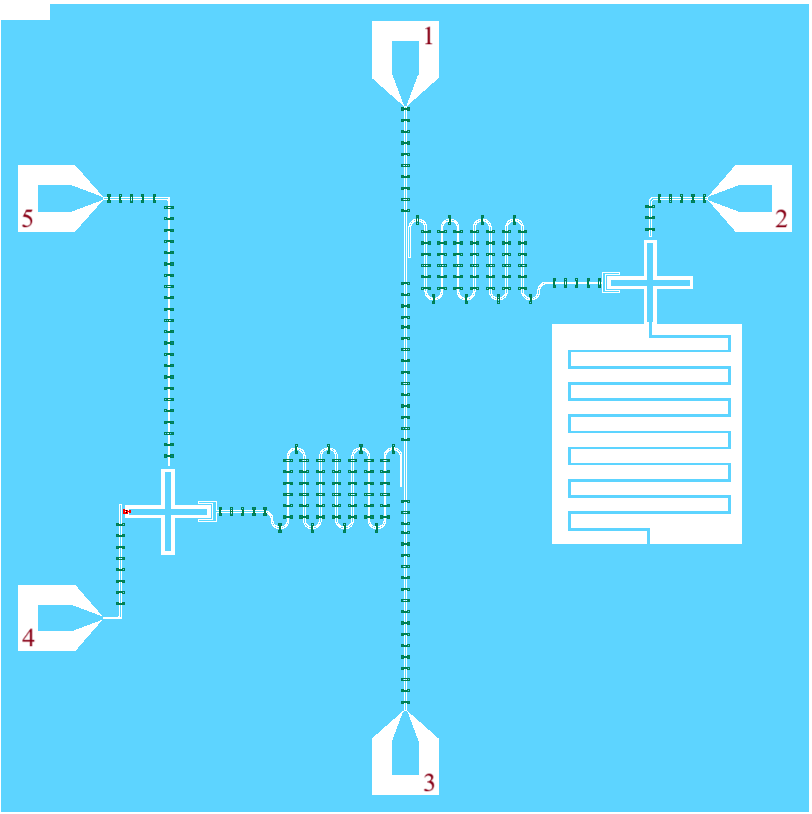}\caption{\label{FullFig}
Design for a 5$\times$5 mm chip: etched 0.25$\mu$m thick aluminum layer atop a 300$\mu$m thick silicon substrate giving effective dielectric constant~\cite{Crowe:2024} $\varepsilon\approx\frac12(\varepsilon_{substrate}+1)=6.225$.} 
\end{figure}
%
%
%%%%%%%%%%%%

The qubit capacitors are made of cross-shaped aluminum arms 240 $\mu$m long and 30 $\mu$m wide, with 30 $\mu$m wide etched trenches. On the RHS we have the meander inductor based qubit, with the following meander inductor dimensions: 6 turns, a 94 $\mu$m lead length, a 10 $\mu$m aluminum strip width, a 90 $\mu$m width between two strips, and a 0.99 mm strip length. As for the transmon qubit (LHS), we have an asymmetric SQUID constructed with Al/Al-Ox/Al tri-layer Josephson Junctions~\cite{Crowe:2024} that are 0.475 and 0.57 $\mu$m wide.

The two coplanar waveguide readout resonators are designed to have resonant frequencies different from one another, and both far from the excitation energies of the qubits~\cite{Crowe:2024}. Their striplines are 10 $\mu$m wide, with 6 $\mu$m trench widths (etched on either side). Aluminum air bridges ensure the grounding is uniform on opposite sides of the coplanar waveguide.

%%%%%%%%%%%%%%%%%%%%%%%%%%%%%%%%%%%%%%%%%%%%%%%%%%%%%%%%%%%%%%%%%%%%%%%%%%%%%%%%%%%%%%%%%%%%
\subsection{Meander inductor based qubit}
\label{MeanderAnalysis}

We consider the meander based qubit shown in~\rf{ZoominRHS}, deriving its anharmonicity based on the following simulation results using InductEx~\cite{Fourie:2017}. The capacitance between the qubit and the ground was found to be $C_{q1}=118.1$ fF. % 18.97 x 6.225 effective perimittivity 
As for the meander inductor, we obtain an inductance of $L_M=18.2$ nH, larger than the semi-analytical prediction of 7.0 nH from Eq.\,(11) of~\cite{Stojanovic2004}. The resulting  ratio of charging energies of the capacitor ($E_{c1}=e^2/2C_{q1}$) and meander inductor ($E_l=\hbar^2 /4e^2L_{M}$) is $E_l/E_{c1}=54.9$.

To obtain the anharmonicity, we take the Hamiltonian~\req{PhiEffAnharm}, with the effective Josephson phase operator $\frac12\hat\varphi_{eff}^2=1-\widehat{\cos{\varphi}}$. Setting $\theta$ and $V_d(t)$ to zero, the nonlinearity in the Hamiltonian can be analyzed in the same manner as in the transmon qubit case (see~\cite{Blais_2004, Koch_2007, Roth:2021lrz} and references therein).

We compute the matrix elements for $\left<m|\widehat{\cos{\varphi}}|n\right>$ in the charge basis following~\cite{Roth:2021lrz,Crowe:2023kle}. Given an analogous Hamiltonian for a pendulum, where $\hat n$ acts as angular momentum and $\hat \phi$ as angular position, we recognize $\left<m|\varphi\right>=e^{i m\varphi}$ and as a result $\left<m|\widehat{\cos{\varphi}}|n\right>=(\delta_{m,n+1}+\delta_{m,n-1})/2$. Combining this contribution with the diagonal components ($E_c$ dependent part of~\req{PhiEffAnharm}), we truncate the matrix to $|n|,|m|\leq100$ Cooper pairs and numerically diagonalize via {\it np.linealg.eig}~\cite{Crowe:2024}. The resulting anharmonicity between the lowest levels
\begin{equation}
\label{anharmNum}
\alpha/h=E_{12}/h-E_{01}/h=187\;{\rm MHz}
\;,
\end{equation}
where $E_{nm}=E_m-E_n$, and the first level excitation energy $E_{01}/h=3.26$ GHz.

A time-independent perturbative approximation for the anharmonicity is also possible~\cite{Crowe:2024}. Applying the nonlinear contributions to the Hamiltonian ($\widehat{\cos{\varphi}}$ expanded to fourth order in the Josephson phase operator $\hat \varphi$) as perturbations to eigenstates of the linear Hamiltonian, we obtain $|\alpha_{pert}|/h=E_{C1}/h=164$ MHz.

To measure the qubit state we apply a readout resonator with a simulated inductance of $L_{r1}=2.40$ nH and capacitance of $C_{r1}=836$ fF. % 134 fF Cg-Cr1 capacitance matrix x 6.225 effective perimittivity 
This gives a quarter-wave resonance frequency of $f_{r1}=1/4\sqrt{LC}=5.58$ GHz (see Eq.\,(4.29) of~\cite{2008PhDTG}, using total inductance and capitance in place of their values per unit length). In comparison, an analytical estimate~\cite{Crowe:2023kle} considering the resonator length $l=4619\;\mu$m gives a frequency of $f_{r1}^{\rm analytic}=c/4l\sqrt{\varepsilon_{eff}}=6.51$ GHz, recalling the effective dielectric constant of the silicon substrate $\varepsilon_{eff}\approx6.225$.

To determine the dispersive shift, we compute the $g$-factor governing the qubit's coupling strength to the readout resonator~\cite{Koch_2007}: 
% $g=2e\beta V_{rms}\left<0|\hat n|1\right>$, where $V_{rms}=\sqrt{hf_{r1}/2C_{r1}}$.  $\beta=C_{q1-r1}/(C_{q1-r1}+C_{q1})$, 
\begin{equation}
\label{gfactor}
g=2e\frac{C_{q1-r1}}{C_{q1-r1}+C_{q1}} \sqrt{\frac{hf_{r1}}{2C_{r1}}} \left<0|\hat n|1\right>
\;,
\end{equation}
where the capacitance between the qubit and readout resonator $C_{q1-r1}=3.08$ fF. Here $\left|0\right>,\left|1\right>$ label the 0th and 1st anharmonic oscillator energy eigenvectors (not to be confused with the charge eigenstates), recalling $\hat n$ is the charge (Cooper pair) number operator. We obtain $g=20.4$ MHz using $f_{r1}$ and $g=22.0$ MHz with $f_{r1}^{\rm analytic}$. The dispersive shift 
\begin{equation}
\label{dispersChi}
\chi=\frac{g^2}{2\pi(f_{r1}-E_{01}/h)}
\;,
\end{equation}
with $\chi=28.5$ kHz using $f_{r1}$, compared with  $\chi=23.7$ kHz using $f_{r1}^{\rm analytic}$.
%
%~\cite{Blais_2007, Gu_2017}

%%%%%%%%%%%%%%%%%%%%%%%%%%%%%%%%%%%%%
%
\begin{figure}[h]
\center
\includegraphics[width=0.8\columnwidth]{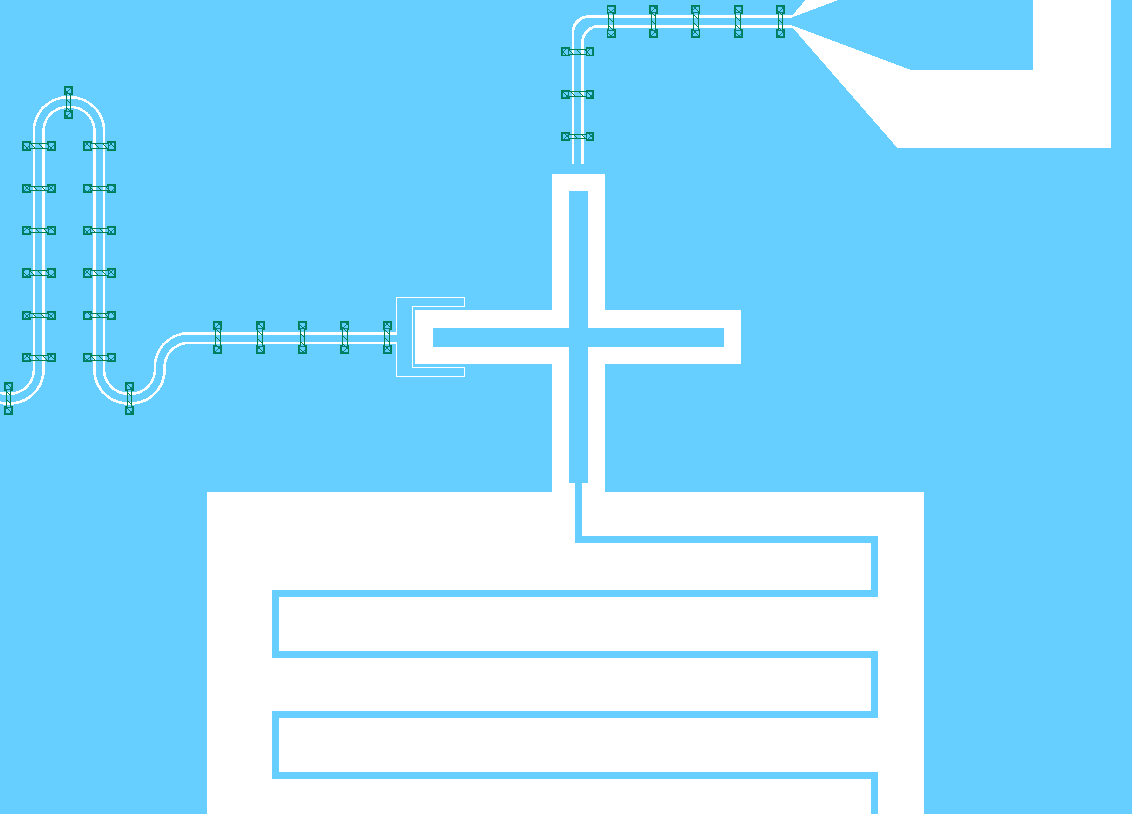}
\caption{\label{ZoominRHS}
Zoom-in to the RHS of \rf{FullFig}: meander inductor based qubit.} 
\end{figure}
%
%
%%%%%%%%%%%%

%%%%%%%%%%%%%%%%%%%%%%%%%%%%%%%%%%%%%%%%%%%%%%%%%%%%%%%%%%%%%%%%%%%%%%%%%%%%%%%%%%%%%%%%%%%%
\subsection{Transmon Qubit}

For comparison, we perform an InductEx simulation of the  transmon qubit shown in~\rf{ZoominLHS}. The capacitance between the transmon qubit and the ground was found to be $C_{q2}=123.8$ fF. The critical current of the SQUID is $I_{c}=31.3$ nA, giving a Josephson inductance of $10.5$ nH. Following the above steps, the junction to capacitor charging energy ratio $E_j/E_{c2}=99.5$. The anharmonicity $\alpha/h=171.5$ MHz, and the excitation energy $E_{01}/h=4.25$ GHz.

%%%%%%%%%%%%%%%%%%%%%%%%%%%%%%%%%%%%%
%
\begin{figure}[h]
\center
\includegraphics[width=0.8\columnwidth]{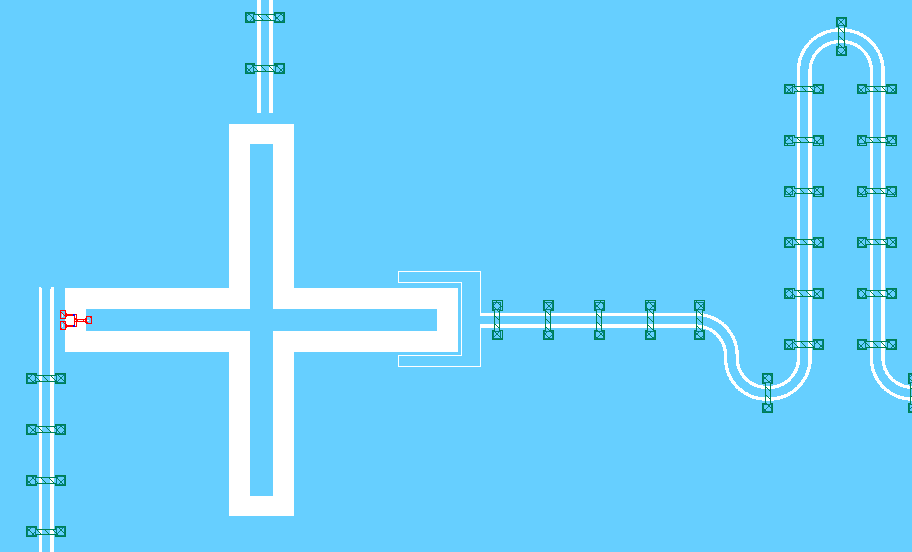}
\caption{\label{ZoominLHS}
Zoom-in to the LHS of \rf{FullFig}: transmon qubit with an asymmetric SQUID.} 
\end{figure}
%
%
%%%%%%%%%%%%

As for the readout resonator, we obtain an inductance of $L_{r2}=2.364$ nH and capacitance of $C_{r2}=828$ fF. The resulting (quarter wave) resonance frequency is $f_{r2}=5.65$ GHz, while the resonator length $l=4575\;\mu$m gives an analytical estimate of $f_{r2}^{\rm analytic}=6.57$ GHz. We find for the capacitance between the qubit and readout resonator $C_{q2-r2}=3.14$ fF, which give a coupling of $g=25.4$ MHz using ~\req{gfactor}, and a dispersive shift of $\chi=44.1$ kHz.

%%%%%%%%%%%%%%%%%%%%%%%%%%%%%%%%%%%%%%%%%%%%%%%%%%%%%%%%%%%%%%%%%%%%%%%%%%%%%%%%%%%%%%%%%%%%
\subsection{Results} \label{sec:RD}
     
% Maybe this moves to conclusion
Combining the above results into\,~\rt{resultsAnharm}, we find that a similar anharmonicity can be achieved comparing the meander inductor based and transmon qubit designs. 

\begin{table}[h]
\begin{center}
 \begin{tabular}{ | l | c | c | c | c | c | c | }
 \hline  
  &       $C_{q}$(fF)      &     $L$(nH)       &    $E_{01}/h$(GHz)   &  $\alpha$(MHz)        &     $g$(MHz)      &    $\chi$(kHz)    
\\ \hline 
meander  &  118  &  18.2  &  3.26  &  187  &  22 & 23.7 % note 23.8 on patent, slightly different resonance frequency
\\ \hline 
JJ  &  124  &  10.5  &  4.25  &  172  &  25.4 & 44.1
  \\ \hline 
% 
%
% R(meander)  &   &   &   &   &   &
%  \\ \hline 
% % 
% %
% R(JJ)  &   &   &   &   &   &
% \\ \hline 
%
%
 \end{tabular}
 \caption{The top row (meander inductor based qubit) shows the qubit capacitance, inductance of the meander inductor, qubit excitation energy, anharmonicity, $g$-factor~\req{gfactor}, and dispersive shift~\req{dispersChi}. The bottom row (standard transmon qubit) shows the same, with the Josephson inductance in place of that of the meander inductor. Values shown are for the choice of $\frac{1}{2}\hat{\varphi}_{eff}^2=1-\widehat{\cos{\varphi}}$. \label{resultsAnharm}}
\end{center}
\end{table}

% Simulated vs heuristic meander inductance comparison 
For completeness, we recall that the meander based qubit calculations use the Inductex simulation result for the inductance of the meander inductor. Had we instead relied on the heuristic formula for the inductance, Eq.\,(11) of~\cite{Stojanovic2004}, we would implement the smaller $L_M=7.0$ nH as opposed to the simulated $L_M=18.2$ nH. In this case we would have: a charging ratio of $E_{l}/E_{c1}=142.4$, anharmonicity $\alpha/h=176.7$ MHz, excitation energy $E_{01}/h=5.37$ GHz, and (using $f_{r1}^{\rm analytic}$) coupling strength $g=28.2$ MHz along with dispersive shift $\chi=110.8$ kHz.

In large part, only the dispersive shift and excitation energy are impacted by the factor $\sim2.6$ difference in the two inductance values. We also recall from~\rs{MeanderAnalysis} differences between simulated results and analytical models for the resonance frequencies of the readout resonators. Most important, however, is the form of the effective phase variable $\hat\varphi_{eff}$ governing the anharmonicity, which will be probed experimentally. Once this is done, we can adjust the dimensions of the meander inductor and readout resonator to optimize qubit performance. 

\section{Conclusion}

In conclusion, we have applied polymer quantization to several superconducting circuits, including the transmon qubit, transmission line resonators, and LC oscillators. Within this framework, the charge offset can be understood as a quantization ambiguity, and more generally, LC circuits generally have anharmonicity due to the discrete spectrum of the number operator. Depending on choices of effective phase variables, we are able to make workable qubit designs with characteristics that are comparable with ordinary transmons. Building on these considerations, we have proposed a specific superconducting qubit design that uses a meander inductor instead of a Josephson junction, thereby avoiding the associated noise sources and fabrication challenges. Given a choice for the effective phase we computed the anharmonicity, dispersive shift, and frequency of this device following from the polymer analysis, demonstrating that the formalism can generate experimentally relevant quantities. 
% We also recall that the meander inductor based qubit results, summarized in~\rs{sec:RD}, depend on whether we use the InductEx simulation or heuristic result for meander inductance. In large part, only the dispersive shift and excitation energy are impacted by the factor $\sim2.5$ difference in the two inductance values. More important is the form of the effective phase variable $\hat\varphi_{eff}$ governing the anharmonicity, which we will measure experimentally. Once this is done, we can adjust the dimensions of the meander inductor to optimize the meander based qubit performance. 

Looking ahead, these results suggest that polymer quantization provides a mathematically consistent framework for describing superconducting circuits, and also a practical tool for informing new designs. The proposed qubit design motivates experimental efforts to probe the extent to which polymer-inspired models can be realized in practice and whether their predicted advantages are realizable. More broadly, the application of techniques originating in quantum gravity to superconducting circuits opens an avenue for cross-fertilization.

\section*{Acknowledgements} \label{sec:acknowledgements}

This work was supported by the NIWC PAC In-house Innovation Program, project number 247. We thank Martin Bojowald for his comments on this manuscript.

% \appendix*
% \input{sections/appendix1.tex}

\bibliographystyle{apsrev4-2}
\bibliography{ref}

\end{document}